\documentclass[twocolumn,superscriptaddress,prl]{revtex4}

\usepackage[T1]{fontenc} 

\usepackage[dvips]{graphicx}
\usepackage{delarray}
\usepackage{amsmath,amssymb,array}
\usepackage{bm}
\usepackage[dvipsnames]{xcolor}

\newcommand{\ket}[1]{| { #1} \rangle}
\newcommand{\bra}[1]{ \langle {#1} |}

\begin{document}
\title{Second law of black hole thermodynamics}

\author{Koji Azuma}
\email{koji.azuma.ez@hco.ntt.co.jp}
\affiliation{NTT Basic Research Laboratories, NTT Corporation, 3-1 Morinosato Wakamiya, Atsugi, Kanagawa 243-0198, Japan}
\affiliation{NTT Research Center for Theoretical Quantum Physics, NTT Corporation, 3-1 Morinosato-Wakamiya, Atsugi, Kanagawa 243-0198, Japan}

\author{Go Kato}
\affiliation{NTT Communication Science Laboratories, NTT Corporation, 3-1 Morinosato Wakamiya, Atsugi, Kanagawa 243-0198, Japan}
\affiliation{NTT Research Center for Theoretical Quantum Physics, NTT Corporation, 3-1 Morinosato-Wakamiya, Atsugi, Kanagawa 243-0198, Japan}

\date{\today}

\begin{abstract}
If simple entropy in the Bekenstein-Hawking area law for a Schwarzschild black hole is replaced with `negative' quantum conditional entropy---which quantifies quantum entanglement---of positive-energy particles of the black hole relative to its outside, a paradox with the original pair-creation picture of Hawking radiation, the first law for black hole mechanics and quantum mechanics is resolved.
However, there was no way to judge experimentally which area law is indeed adopted by black holes. 
Here, with the no-hair conjecture, we derive the perfect picture of a second law of black hole thermodynamics for any black hole from the modified area law, rather than Bekenstein's generalized one from the original area law.
The second law is testable with an event horizon telescope, in contrast to Bekenstein's. 
If this is confirmed, the modified area law could be exalted to the first example of fundamental equations in physics which cannot be described without the concept of quantum information. 
\end{abstract}
\maketitle


A black hole is one of most beautiful but mysterious objects in our universe. Although its carrier started merely as a purely theoretical object in solutions of the Einstein equation in general relativity, nowadays, it is a target of observational astrophysics \cite{E19-1,E19-2,E19-3,E19-4,E19-5,E19-6,A16}.
Apparently, 
its typical picture that black holes absorb only and nothing can escape from them looked highly irreversible, compared with normal stars. However, this is merely a view for black holes in the regime of classical general relativity and not the case for the quantum world. In particular, remarkably, Hawking has developed a semi-classical picture \cite{H74,H75} where thermal radiation occurs from a Schwarzschild black hole---although it is regarded as `useless' classically because we cannot distil energy from it, in contrast to Kerr or charged black holes \cite{P69,MTW}. As a result, Hawking famously described it as `{\it a ``black hole'' is not completely black}' \cite{H76}. However, this Hawking radiation gave us more serious puzzles about the consistency between such black hole mechanics and quantum mechanics.

A puzzle appears \cite{AS18,BPZ13} when we combine the Hawking radiation with the first law of black hole mechanics \cite{BCH73}, the Bekenstein-Hawking equation \cite{B73,B74,H74,H75,H76} and quantum mechanics. The first law of black hole mechanics is associated with the energy conservation law: for a stationary black hole $B$, in Planck unit, we have
\begin{equation}
\label{eq:firstlaw}
{\rm d}M_B=\frac{\kappa_B}{8\pi}{\rm d}A_B + \Omega_B {\rm d}J_B + \phi_B {\rm d}Q_B, 
\end{equation}
where $M_B$ is the mass, $A_B$ is the area of the event horizon, $\kappa_B$ is the surface gravity, $J_B$ is the angular momentum, $\Omega_B$ is the angular velocity, $Q_B$ is the charge and $\phi_B$ is the electrostatic potential of the black hole. 
Here $\Omega_B {\rm d}J_B + \phi_B {\rm d}Q_B$ in the first law corresponds to the change of black hole energy as work, and thus, in general,
\begin{equation}
\delta {\cal Q}_B := {\rm d}M_B- \Omega_B {\rm d}J_B - \phi_B {\rm d}Q_B
\end{equation}
is deemed to be the change of heat.
On the other hand,
the Bekenstein-Hawking equation is an area law for the black hole:
\begin{equation}
\frac{{\rm d}A_B}{4}={\rm d} S(B), \label{eq:Beken}
\end{equation}
where $S(B)$ is the entropy of the black hole $B$.
However, those laws (\ref{eq:firstlaw}) and (\ref{eq:Beken}) are inconsistent \cite{AS18,BPZ13} with the pair-creation picture of Hawking radiation, in a quantum mechanical point of view. More precisely, Hawking's finding is that an observer at the future infinity receives thermal radiation $H^+$ with Hawking temperature $\kappa_B/(2 \pi)$ from a Schwarzschild black hole (with $\Omega_B=0$ and $\phi_B=0$), whose purification partner $H^-$ is regarded as having negative energy and tunnelling into the black hole $B$. The fact that the observer receives the positive energy of the radiation $H^+$ means ${\rm d} M_B<0$ from the energy-conservation law, implying ${\rm d}A_B <0$ according to the first law (\ref{eq:firstlaw}) with $\Omega_B=0$ and $\phi_B=0$. On the other hand, quantum mechanics tells us that the purification partner $H^-$ has the same {\it positive} entropy as the thermal radiation [$S(H^-)=S(H^+)>0$ with any unitary-invariant entropic measure $S$ (e.g., the von Neumann entropy) because $H^+H^-$ is in a pure state]. Thus, the fact that the black hole receives this purification partner $H^-$ with {\it positive} entropy means ${\rm d}S(B) >0$, implying ${\rm d}A_B >0$ according to the Bekenstein-Hawking equation (\ref{eq:Beken}). Hence, Hawking's original picture for Hawking radiation leads to a paradox about the direction of the change of the area $A_B$.

Recently, it has been argued \cite{AS18} that this paradox is resolved if we assume that a black hole stores quantum entanglement, rather than simple entropy, i.e., if the Bekenstein-Hawking equation (\ref{eq:Beken}) is modified as
\begin{equation}
\frac{{\rm d}A_B}{4} = {\rm d}I(\bar{B} \rangle B^+):=-{\rm d}S(\bar{B}|B^+)={\rm d}S(B^+)-{\rm d}S(B^-)\label{eq:m1}
\end{equation}
for a stationary Schwarzschild black hole $B$,
where $I(X \rangle Y)$ is called the coherent information \cite{SN96,HOW05,HOW07} from $X$ to $Y$ and $S (X|Y)$ is the conditional entropy defined by $S(X|Y):=S(XY)-S(Y)$ with the von Neumann entropy $S(X):=S(\hat{\rho}_X):=-{\rm Tr}[\hat{\rho}_X\ln \hat{\rho}_X]$ for a system $X$ in a state $\hat{\rho}_X$. 
The coherent information is positive only in the quantum world, and is indeed associated with one-way distillable entanglement in quantum information theory.
In the modified area law~(\ref{eq:m1}),
if it applies to a Schwarzschild black hole \cite{AS18}, we assume the followings: (i) the black hole $B$ is composed not only of normal positive-energy particles $B^+$ (like particles having fallen down into the black hole from spatial infinity), but also of negative-energy particles $B^-$ generated in its inside by the Hawking process (like the Hawking particles $H^-$ whose appearance decreases the area $A_B$), i.e. $B=B^+B^-$ (that is, ${\cal H}_B = {\cal H}_{B^+} \otimes {\cal H}_{B^-}$, where ${\cal H}_X$ is the Hilbert space for a system $X$); (ii) the whole system $B\bar{B}=B^+B^-\bar{B}$, by including a system $\bar{B}$ in the outside of the black hole $B$, can always be in a pure state; (iii) the free evolution of the black hole $B$ is described by a unitary operator in the form $\hat{U}_{B^+} \otimes \hat{V}_{B^-}$. 
Notice that Eq.~(\ref{eq:m1}) is equivalent to the Bekenstein-Hawking equation (\ref{eq:Beken}) for any process with ${\rm d}S(B^-)=0$, which reproduces \cite{AS18} all known results shown with the original equation (\ref{eq:Beken}). But, Eq.~(\ref{eq:m1}) not only solves the paradox above in contrast to Eq.~(\ref{eq:Beken}), but also is free \cite{AS18} from other paradoxes, such as the information loss paradox \cite{P92} and the firewall paradox \cite{AMPS13}.

In this paper, we revisit thermodynamics of black holes, from the view of quantum information.
In particular, we first show that the modified area law (\ref{eq:m1}) holds not only for stationary Schwarzschild black holes but also for any stationary black hole following the first law (\ref{eq:firstlaw}), by reinterpreting $B^+$ as a positive-heat part of the black hole and $B^-$ as its negative-heat part. Then, with this generalized area law (\ref{eq:m1}), the no-hair conjecture and Patovi's model for a thermal bath \cite{P89,Peres}, we derive a second law of arbitrary thermodynamic process~$\alpha'$ which converts a black hole $B$ from a stationary state $B_1$ into a stationary state $B_2$ by interacting with a thermal bath $R$:
\begin{equation}
 - \int_{B_1 \xrightarrow{\alpha'}  B_2} \beta_R \delta  {\cal Q}_R  \le 
 \frac{A_{B_2}}{4} -\frac{A_{B_1}}{4} , \label{eq:main-heat}
\end{equation}
where $\beta^{-1}_R$ and $\delta {\cal Q}_R$ represent the `redshifted' temperature and the received heat of the thermal bath $R$, which is related with hole's received heat ${\cal Q}_B$ as $\delta {\cal Q}_R=- \delta {\cal Q}_B$.
Note that the area of the black hole in this second law is analogous to the entropy of a normal thermodynamic system in the second law of thermodynamics.
Also notice that all the quantities in the second law (\ref{eq:main-heat}) are all observables. Therefore, the law is testable, for instance, by using an event horizon telescope \cite{E19-1,E19-2,E19-3,E19-4,E19-5,E19-6,GJC12}, in contrast to Bekenstein's generalized second law \cite{B74} associated with Eq.~(\ref{eq:Beken}) (although the modified area law (\ref{eq:m1}) also follows Bekenstein's generalized second law \cite{AS18}):
\begin{equation}
\frac{{\rm d}A_B}{4} +{\rm d} S(\bar{B}) \ge 0, \label{eq:B2}
\end{equation}
where the entropy $S(\bar{B})$ of the outside $\bar{B}$---which is not a direct macroscopic observable along a process beyond a quasi-static one---is included.

{\it Entropy conservation and trajectory.}---Let us start by explaining how to evaluate the right-hand side of Eq.~(\ref{eq:m1}) for a black hole $B=B^+B^-$ composed of the positive-heat part $B^+$ and the negative-heat part $B^-$. This can be done by using the following rule from the entropy conservation law in quantum mechanics:
if a stationary black hole $B=B^+B^-$ stores an infalling (or tunnelling \cite{H75}) object $C$---whose purification partner belongs to the outside $\bar{B}$---having positive (negative) `heat' ${\cal Q}_C:=E_C-\Omega_B l_C-\phi_B Q_C\ge0$ (${\cal Q}_C \le 0$) with energy-at-infinity $E_C$ \cite{MTW}, axial component $l_C$ of angular momentum and charge $Q_C$---evaluated at event of crossing, then the entropy changes $\Delta S(B^\pm):=S(B'^\pm)-S(B^\pm)$ of the positive-heat and negative-heat parts of the black hole are $\Delta S(B^+)=S(C)$ and $\Delta S(B^-)=0$ ($\Delta S(B^-)=S(C)$ and $\Delta S(B^+)=0$), where $S(C)$ is the von Neumann entropy of the system $C$ at that event, and $B'=B'^+ B'^-$ is the black hole after capturing the object $C$ in its positive-heat (negative-heat) part at the vicinity of the horizon and then evolving unitarily in each part according to a unitary operator $\hat{U}_{B'^+} \otimes \hat{V}_{B'^-}$, i.e., ${\cal H}_{B'^+}={\cal H}_{B^+} \otimes {\cal H}_C$ and ${\cal H}_{B'^-}= {\cal H}_{B^-}$ (${\cal H}_{B'^-}={\cal H}_{B^-} \otimes {\cal H}_C$ and ${\cal H}_{B'^+}= {\cal H}_{B^+}$). Here, for each case, $\Delta S(B^\pm)=S(C)$ and $\Delta S(B^\mp)=0$ follow from $S(B'^\pm) =S(B^\pm C)=S(B^\pm)+S(C) $ and $S(B'^\mp)=S(B^\mp)$, where we have used the invariance of the von Neumann entropy under any unitary operation (i.e., $S(B'^\pm) =S(B^\pm C)$ and $S(B'^\mp)=S(B^\mp)$ under the unitary $\hat{U}_{B'^+} \otimes \hat{V}_{B'^-}$) and $I(B^\pm:C)=0$ from the condition for the purification partner of system $C$.

In the rule, it is assumed that object $C$ is very small, that is, its size and mass are much smaller than those of the hole and it has sufficiently small charge, so that its gravitational/electromagnetic radiation is negligible. Thus, it moves very nearly along a test-particle trajectory, which approaches the horizon with future-pointing 4-momentum when ${\cal Q}_C \ge 0$ while with past-pointing 4-momentum when ${\cal Q}_C \le 0$ (e.g., see Sec.~33.7 of Ref.~\cite{MTW}).
Thus, the negative-heat object $C$ in the rule can be composed only of the purification partner $H^-$ of the radiation $H^+$, generated in the {\it inside} of the hole $B$ by Hawking radiation as a quantum effect,
because there is no test particle in the outside $\bar{B}$ whose orbit can cross the horizon with negative heat ${\cal Q}_C <0$. Energy-at-infinity
$E_C$, angular momentum $l_C$, charge $Q_C$ and von Neumann entropy $S(C)$ in the above rule should be determined once we are given a density operator $\hat{\rho}_C$ which is the quantum description of the internal state of system $C$ at event of crossing. This is similar treatment made in Ref.~\cite{B73}.
Therefore, in this framework, the rule for entropy is analogous to the energy conservation law in general relativity (e.g., see Sec.~33.7 of Ref.~\cite{MTW}).

{\it Quasi-static process.}---To derive our second law~(5), we first show that the modified area law~(\ref{eq:m1}) holds not only for Schwarzschild black holes but also for arbitrary black holes with Hawking radiation, by generalizing the formulation in Ref.~\cite{AS18}. 
This enables us to introduce the concept of a quasi-static process for any stationary black hole and to conclude the coherent information as a state of quantity from the no-hair conjecture.

Let us consider a stationary black hole $B$, which emits a Hawking pair $H^+H^-$ in state 
\begin{align}
\ket{\chi}_{H^+ H^-} :=& \exp[r_{\omega'}(\hat{a}_k^\dag \hat{b}_{-k}^\dag- \hat{a}_k\hat{b}_{-k})]\ket{{\rm vac}} \nonumber \\
= & \frac{1}{\cosh r_{\omega'}} \sum_{n=0}^{\infty} \tanh^n r_{\omega'} \ket{n}_{H^+} \ket{n}_{H^-} ,\label{eq:two-mode}
\end{align}
where $\hat{a}_k$ and $\hat{b}_{-k}$ are annihilation operators associated with the positive-heat particles $H^+$ and the negative-heat particles $H^-$ respectively, the parameter $r_{\omega'}$ is related \cite{H76} to a mode with frequency $\omega$, angular momentum $m$ about the axis of rotation of the black hole, and charge $e$ via 
$\omega':=\omega-m \Omega_B -e \phi_B$
and
$\exp(-\pi \omega'/\kappa_B)=\tanh r_{\omega'}$, 
and the effective mode frequency $\omega'$ will follow some dispersion relation $\omega'=\omega'(\pm k)$. The free Hamiltonian of particles $H^\pm$ is $\pm \omega' \hat{n}_{H^\pm}$ \cite{H96}, where 
$\hat{n}_{{ H}^+}:=\hat{a}_k^\dag \hat{a}_k$ and $\hat{n}_{{ H}^-}:= \hat{b}_{-k}^\dag \hat{b}_{-k}$.
In the pair creation picture, the negative-heat particles ${H}^-$  appear in a mode falling into the black hole (i.e. on a worldline crossing the event horizon), while the positive-heat particles ${ H}^+$ appear in a mode propagating from the vicinity of the event horizon to a static observer at infinity. 
The reduced state of the positive-heat particles ${ H}^+$ is the Gibbs state with temperature~$\beta_{ H}^{-1}=\kappa_B/(2\pi)$,
\begin{align}
\hat{\chi}_{{ H}^+}:=& {\rm Tr}_{H^-} [\ket{\chi}\bra{\chi}_{H^+H^-}]   \nonumber \\
=&\frac{1}{\cosh^2 r_{\omega'}} \sum_{n=0}^\infty \tanh^{2n}r_{\omega'} \ket{n}\bra{n}_{{H}^+} \nonumber \\
=& \frac{e^{-\beta_{H} \omega' \hat{n}_{{H}^+} }}{Z_{\beta_{H} }}. \label{eq:chi}
\end{align}
Since this satisfies
$
- \ln \hat{\chi}_{{ H}^+}= \beta_{ H}  \omega' \hat{n}_{{ H}^+} +\ln Z_{\beta_{ H}} \hat{1}_{{ H}^+}
$
for the partition function $Z_{\beta_{ H}}:=(1-e^{-\frac{2\pi \omega'}{\kappa_B}})^{-1}=(1-e^{-\beta_{ H} \omega'})^{-1}$, we have
\begin{equation}
\label{eq:Gibbs_S}
S({ H}^+)=\beta_{ H}  \omega' n_{{ H}^+} + \ln Z_{\beta_{ H}},
\end{equation}
where
\begin{equation}
n_{{ H}^+} ={\rm Tr}[ \hat{n}_{ H^+}  \hat{\chi}_{{ H}^+} ] = \frac{1}{e^{\beta_{ H}\omega'}-1} .\label{eq:nH+}
\end{equation}
Hence, the positive-heat particles satisfy
\begin{align}
\frac{1}{\omega'} \frac{{\rm d} S( H^+) }{ {\rm d} n_{{ H}^+}}
=&\frac{1}{\omega'} \biggl( \beta_{ H} \omega' + \omega' n_{{ H}^+  } \frac{{\rm d} \beta_{ H}} {{\rm d} n_{{ H}^+}} \nonumber \\
 &+ \frac{1}{Z_{\beta_{ H}}} \frac{\partial Z_{\beta_{ H}}}{\partial \beta_{ H}} \frac{{\rm d} \beta_{ H}} {{\rm d} n_{{ H}^+}} \biggr) 
=\beta_{ H} \label{eq:deriv}
\end{align}
for given $\omega'$. Therefore, the emission of positive-heat particles $H^+$ from the event horizon is pure thermal radiation at the Hawking temperature $\beta_{ H}^{-1}=\kappa_B/(2\pi)$. 
Note that Eq.~(\ref{eq:deriv}) implies that $\omega' {\rm d} n_{H^+}$ is the received heat $\delta {\cal Q}_{H^+}$ of positive-heat particles $H^+$ in any quasi-static process (i.e., $\omega' {\rm d} n_{H^+}=\delta {\cal Q}_{H^+}$), because $H^+$ is initially in a thermal equilibrium state and $\beta_{ H}^{-1}{\rm d} S( H^+)= \delta {\cal Q}_{H^+}$ holds for the process.

Now, we consider a process where the black hole emits the Hawking radiation $H^+$ to infinity, while, from the outside, it absorbs an infalling thermal bosonic system $C$ with the same effective mode frequency $\omega'$, positive heat ${\cal Q}_C(\ge 0)$ and entropy $S(C)$. Here the system $C$ is initially decoupled with the black hole $B$ before this absorption.
In this process, the black hole $B$ loses heat ${\cal Q}_{H^+}$ of positive-heat particles $H^+$ but receives entropy $S(H^-)$ of negative-heat particles $H^-$ through the Hawking radiation, while it receives positive heat ${\cal Q}_C$ and entropy $S(C)$ by absorbing such normal (positive-heat) particles $C$. Therefore, in this process, the heat change $\Delta {\cal Q}_B$ of the black hole $B$ and the change $\Delta I(\bar{B}\rangle B^+)$ of coherent information are given by
\begin{align}
&\Delta {\cal Q}_B = {\cal Q}_C -{\cal Q}_{H^+}, \label{eq:ch-e} \\
&\Delta I(\bar{B}\rangle B^+) = S(C) -  S(H^-) = S(C) -  S(H^+), \label{eq:ch-I}
\end{align}
where we have used conservation laws for energy, charge, and axial component of angular momentum (see Sec.~33.7 of Ref.~\cite{MTW}) in Eq.~(\ref{eq:ch-e}), and
we have used $S(H^-)=S(H^+)$ for the pure state $\ket{\chi}_{H^+H^-}$ in Eq.~(\ref{eq:ch-I}).

Suppose that the system $C$ is in a thermal state $\hat{\chi}_C$ as in Eq.~(\ref{eq:chi}), with the Hawking temperature $\beta_{ H}^{-1}=\kappa_B/(2\pi)$.
Then, we have 
\begin{equation}
\omega' n_C=\omega' n_{H^+}, \label{eq:eq}
\end{equation}
meaning ${\cal Q}_C={\cal Q}_{H^+}$ from Eq.~(\ref{eq:nH+})
and $S(H^+)=S(C)$ from Eq.~(\ref{eq:Gibbs_S}).  Hence, in this case, the above process provides $\Delta {\cal Q}_B=0$ and $\Delta I(\bar{B}\rangle B^+)=0$ from Eqs.~(\ref{eq:ch-e}) and (\ref{eq:ch-I}), which conclude $\Delta A_B=0$, either from the first law (1) for stationary black holes or from the modified area law (4).
Therefore, as long as this equilibrium process is repeated, say if a black hole interacts with thermal systems with the Hawking temperature $\beta_H^{-1}$, the black hole can be exactly in a stationary state. 
This is in contrast to the case for the Bekenstein-Hawking equation (3) (see Ref.~\cite{AS18}).

Let us move on to a case where the above equilibrium process is repeated, but at some point, it deviates slightly from its equilibrium version, accompanied by small changes on the system $C$ and the Hawking radiation $H^+$ such that ${\cal Q}_C: \omega' n_C \to \omega' n_C  + \omega' \Delta n_C$ and ${\cal Q}_{H^+} : \omega' n_{H^+} \to \omega' n_{H^+}  +\omega' \Delta n_{H^+}$. 
For this perturbation, from Eq.~(\ref{eq:eq}), Eq.~(\ref{eq:ch-e}) becomes
\begin{align}
\Delta {\cal Q}_B =& \omega' n_C  + \omega' \Delta n_C -( \omega' n_{H^+}  +\omega' \Delta n_{H^+}) \nonumber \\
=& \omega'  \Delta n_C  - \omega' \Delta n_{H^+}, \label{eq:dE}
\end{align}
while Eq.~(\ref{eq:ch-I}) becomes
\begin{align}
\Delta I(\bar{B}\rangle B^+) =& \beta_{ H}  \omega' n_{C} + \ln Z_{\beta_{ H}}+\Delta S(C) \nonumber \\
&- (\beta_{ H}  \omega' n_{{ H}^+} + \ln Z_{\beta_{ H}}+\Delta S(H^+)) \nonumber \\
=& \Delta S(C) -\Delta S(H^+),  \label{eq:dI}
\end{align}
using Eq.~(\ref{eq:Gibbs_S}).
However, as long as the perturbation is small enough to be regarded as a quasi-static process for system $C$ and Hawking radiation $H^+$, the difference $\Delta X$ on a quantity $X$ can be regarded as its derivative ${\rm d } X$ and Eq.~(\ref{eq:deriv}) should hold. Hence, we have 
\begin{align}
{\rm d}I(\bar{B}\rangle B^+) =& {\rm d} S(C) -{\rm d} S(H^+) \nonumber \\
= &\beta_H ( \omega' {\rm d} n_C - \omega' {\rm d} n_{H^+}) 
= \beta_H {\rm d} {\cal Q}_B  .  
\label{eq:deri}
\end{align}
from Eqs.~(\ref{eq:dE}) and (\ref{eq:dI}).
Combined with the first law (1) for stationary black holes, this concludes Eq.~(\ref{eq:m1}).

If the above perturbation is used as a quasi-static process to change the black hole with keeping it in stationary states, we can integrate Eq.~(\ref{eq:m1}), leading to 
$I(\bar{B}\rangle B^+)=A_B/4+c$
with a constant $c$. Since $A_B$ is a quantity of the state of a black hole according to the no-hair conjecture, this equation shows that
$I(\bar{B}\rangle B^+)$ is also a quantity of the state. This will be used later.

{\it Arbitrary thermodynamic process.}---We introduce a thermodynamic process, perhaps beyond a quasi-static process, 
where a black hole $B$ interacts with a thermal
bath $R$. 
In particular, we consider a process where (1) a thermal bath $R$ falls towards a black hole $B$ from outside $\bar{B}$ with redshifted temperature $\beta_R^{-1}$, energy-at-infinity $E_R$, charge $Q_R$ and the axial component $l_R$ of the angular momentum on black hole's rotation axis, (2) then interacts with the black hole $B$ near the horizon, and (3) finally comes back to spatial infinity.
We assume that the interaction near the horizon between the bath $R$
and the hole $B$ is described by the following two elementary
processes: (a) unitary interaction between positive-heat particles $B^+$ of the black hole $B$ and the thermal bath $R$, and (b) Hawking radiation, which is a unitary interaction between the negative-heat part $B^-$ of the black hole $B$ and a system with an extremely low temperature (like the vacuum) in the bath $R$. 

First, we introduce a thermal bath $R$, which is essentially the same as Partovi's model \cite{P89,Peres}. The thermal bath $R$ is assumed to be composed of a huge number of very small particles $r_i$ in a Gibbs state $\hat{\sigma}_{r_i}$ with redshifted temperature $\beta_{r_i}^{-1}$, that is, $R=\bigotimes_i r_i$. The Gibbs state $\hat{\sigma}_{X}$ for system $X$ near the horizon is 
\begin{equation}
\hat{\sigma}_X = \frac{1}{Z_{\beta_X}}  e^{-\beta_X \hat{\cal Q}_X},
\end{equation}
where 
$
\hat{{\cal Q}}_X:=\hat{H}_X -\Omega_B \hat{l}_X -\phi_B \hat{Q}_X
$
is the observable of heat ${\cal Q}_X$, $\hat{H}_X$ is the Hamiltonian associated with energy-at-infinity $E_X$, $\hat{l}_X$ is the observable of angular momentum $l_X$, and $\hat{Q}_X$ is the observable of charge $Q_X$. For any conversion of a system $X$ from a Gibbs state $\hat{\sigma}_X $ to an arbitrary state $\hat{\rho}_{X'}$, we have
\begin{align}
-\Delta S(X)+\beta_X \Delta {\cal Q}_X 
=&{\rm Tr}[\hat{\rho}_{X'} \ln \hat{\rho}_{X'} +\beta_X \hat{\rho}_{X'}\hat{{\cal Q}}_X ] + \ln Z_{\beta_X} \nonumber \\
=&{\rm Tr}[\hat{\rho}_{X'} (\ln \hat{\rho}_{X'} +\beta_X \hat{\cal Q}_X  + \ln Z_{\beta_X})] \nonumber \\
=& {\rm Tr}[\hat{\rho}_{X'} (\ln \hat{\rho}_{X'} - \ln \hat{\sigma}_X) ] \nonumber\\
=&D(\hat{\rho}_{X'} \| \hat{\sigma}_{X}) \ge 0, \label{eq:1}
\end{align}
where $\Delta  {\cal Q}_X :={\cal Q}_{X'} - {\cal Q}_{X}$ by denoting the expectation value of an observable $\hat{Y}$ as $Y$, and $D(\hat{\rho}_{X'} \| \hat{\sigma}_{X})$ is the relative entropy and is non-negative.

Let us consider the elementary process (a). In this process, the positive-heat part $B^+$ interacts with a system $r_i$, which is described by a unitary interaction $\hat{U}_{B^+ r_i\to B'^+  r_i'}$. Thus we have
\begin{align}
\Delta_i S(B^+) +\Delta_i S(r_i) = & I(B'^+: r'_i)-I(B^+ : r_i) \nonumber \\
=& I(B'^+:r'_i) \ge 0, \label{eq:2}
\end{align}
where we assumed that the particle $r_i$ is initially decoupled with $B^+$, that is, $I(B^+ : r_i)=0$, and $\Delta_i=\Delta$ although $\Delta_i$ has index $i$ to describe that this change is brought by interaction with particle $r_i$. 
Then, since $r_i$ is initially in a Gibbs state which follows Eq.~(\ref{eq:1}), from Eqs.~(\ref{eq:1}) and (\ref{eq:2}), we have
\begin{equation}
\Delta_i S(B^+)+ \beta_{r_i} \Delta_i {\cal Q}_{r_i} \ge 0. 
\end{equation}
Since $\Delta_i S(B^-)=0$
during this process, this concludes
\begin{equation}
\Delta_i I(\bar{B}\rangle B^+) \ge - \beta_{r_i} \Delta_i {\cal Q}_{r_i}.\label{eq:4}
\end{equation}

Next, let us consider the Hawking radiation (b). This process originally occurs between the black hole $B$ and a system $r_j$ in the vacuum state at the outside, by giving them a Hawking pair in a pure entangled state with $S(H^+)=S(H^-)$. Thus, even if we consider Hawking radiation in a more practical scenario where the system $r_j$ in the thermal bath $R$ has an extremely low but nonzero temperature $\beta_{r_j}^{-1}$, 
\begin{align}
\Delta_j S(B^-) = \Delta_j S(r_j) \label{eq:5}
\end{align}
would be a good approximation.
Then, from Eqs.~(\ref{eq:1}) and (\ref{eq:5}), we have
\begin{equation}
-\Delta_j S(B^-) + \beta_{r_j} \Delta_j  {\cal Q}_{r_j}  \ge 0. \label{eq:echange}
\end{equation}
Indeed, a model for the practical Hawking radiation shows that
this inequality itself holds when $\beta_{r_j}^{-1}$ is low enough with $\beta_H^{-1} \ge \beta_{r_j}^{-1}$, although $\Delta_j S(B^-) \ge \Delta_j S(r_j)$ holds (see Appendix).
Also, the inequality (\ref{eq:echange}) implies that $B^-$ has a negative heat spectrum.  In particular, since $\Delta_j {\cal Q}_{B}+\Delta_j {\cal Q}_{r_j}=0$ holds from 
conservation laws for energy, charges, and axial component of angular momentum (e.g., see Sec.~33.7 of Ref.~\cite{MTW}) and $\Delta_j {\cal Q}_{B}= \Delta_j {\cal Q}_{B^+}+\Delta_j{\cal Q}_{B^-} =\Delta_j {\cal Q}_{B^-}$ holds for this process, the inequality (\ref{eq:echange}) means that $\Delta_j S(B^-) \ge 0 $ implies $\Delta_j  {\cal Q}_{B^-}  \le 0$.
Since $\Delta_j S(B^+)=0$ during this process, Eq.~(\ref{eq:echange}) concludes
\begin{equation}
\Delta_j I(\bar{B}\rangle B^+) \ge - \beta_{r_j} \Delta_j {\cal Q}_{r_j}.\label{eq:7}
\end{equation}

From Eqs.~(\ref{eq:4}) and (\ref{eq:7}), for either elementary thermodynamic process (a) or (b), we conclude
\begin{equation}
\Delta_i I(\bar{B}\rangle B^+) \ge - \beta_{r_i}  \Delta_i {\cal Q}_{r_i}     
\end{equation}
for any interaction with a thermal system $r_i$. If we assume that each system $r_i$ is so small that each $\Delta_i$ is regarded as derivative, we can rephrase this inequality as
\begin{equation}
{\rm d} I(\bar{B}\rangle B^+) \ge - \beta_{R}  \delta {\cal Q}_{R}.     \label{eq:main}
\end{equation}
Note that a heat conservation law, $ \delta {\cal Q}_{R}=-  \delta {\cal Q}_{B}$, holds from conservation laws for energy, charge, and axial component of angular momentum (e.g., see Sec.~33.7 of Ref.~\cite{MTW}).

{\it Cycle.}---Let us consider a cyclic process where the black hole $B$ starts from a stationary state $B_1$ and comes back to the initial state. Then, we have
\begin{equation}
\oint {\rm d} I(\bar{B}\rangle B^+) =0, \label{eq:c}
\end{equation}
because the coherent information $I(\bar{B} \rangle B^+)$ is a quantity of the state, stemming from the no-hair conjecture. Let us divide this cycle into two path. The first path is any thermodynamic process $\alpha'$, perhaps beyond a quasi-static process, where the black hole starts from the initial stationary black hole $B_1$ to another stationary black hole $B_2$. The second path is a quasi-static process $\alpha$ which starts from the stationary black hole $B_2$ and comes back to the initial stationary black hole $B_1$. In this path, the black hole satisfies Eq.~(\ref{eq:m1}). Combined with Eqs.~(\ref{eq:main}) and (\ref{eq:c}), this concludes the second law~(5) through
\begin{align}
- \int_{B_1 \xrightarrow{\alpha'}  B_2} \beta_R \delta  {\cal Q}_R  \le &
\int_{B_1 \xrightarrow{\alpha'}  B_2} {\rm d} I(\bar{B}\rangle B^+)  \nonumber \\
=& \oint{\rm d} I(\bar{B}\rangle B^+) 
- \int_{B_2 \xrightarrow{\alpha}  B_1} {\rm d} I(\bar{B}\rangle B^+) \nonumber \\
=& \int_{B_1\xrightarrow{\alpha}  B_2} {\rm d} I(\bar{B}\rangle B^+)  \nonumber \\
=& \frac{1}{4} \int_{B_1\xrightarrow{\alpha}  B_2} {\rm d} A_B.
\end{align}

{\it Discussion.}---Like the second law of normal thermodynamics, our law has many implications. For instance, if $\delta {\cal Q}_B =-\delta {\cal Q}_R \ge 0$ holds---which is valid for any classical test-particle orbits crossing the horizon from the outside $\bar{B}$ (e.g., see Sec.~33.8 of Ref.~\cite{MTW}), 
the second law (\ref{eq:main-heat}) is reduced to Hawking's area theorem ${\rm d} A_{B}\ge  0$---which is derived in the regime of general relativity \cite{H71}. 
Another important implication appears if we consider a stationary black hole $B$ which receives infinitesimally small heat $-\delta {\cal Q}_R$, with keeping its stationarity, i.e., $\delta {\cal Q}_B=4^{-1}\beta_H^{-1} {\rm d} A_B$ from the first law (\ref{eq:firstlaw}). Then, the second law (\ref{eq:main-heat}) and heat conservation law $\delta {\cal Q}_B=-\delta {\cal Q}_R$ imply ${\rm d} A_B \ge - 4 \beta_R \delta {\cal Q}_R = 4 \beta_R \delta {\cal Q}_B = \beta_R \beta_H^{-1} {\rm d}A_B$. Therefore, the black hole $B$ becomes bigger (${\rm d}A_B \ge 0$) when the temperature $\beta^{-1}_H$ of the black hole is less than $\beta_R^{-1}$ ($\beta_H^{-1} \le \beta_R^{-1}$), and it becomes smaller (${\rm d}A_B\le 0$) when its temperature $\beta^{-1}_H$ is higher than $\beta_R^{-1}$ ($\beta_H^{-1} \ge \beta_R^{-1}$).
Many other implications would appear; however, the most important point of this paper is that the second law (\ref{eq:main-heat}) is derived from the equation (\ref{eq:m1}) but cannot from the original one~(\ref{eq:Beken}), and its validity is testable with current technology of observational astrophysics \cite{E19-1,E19-2,E19-3,E19-4,E19-5,E19-6,A16}. This test is needed to answer fundamental questions whether a black hole stores quantum entanglement, as suggested by Eq~(\ref{eq:m1}), and nevertheless, as implied by Eq.~(\ref{eq:main-heat}), whether the black hole is still analogous to a normal thermodynamic system, as Bekenstein and Hawking have originally imagined \cite{B73,B74,H74,H75,H76}.

{\it Acknowledgements.}---We are especially thankful to S.~Subramanian for insightful discussion that has continued since the first paper \cite{AS18} on this project. We also thank T. Honjo, K. Inaba, H.-K. Lo, W. J. Munro, Y. Nakata, K. Shimizu and T. Takayanagi for helpful discussion. K.A. acknowledges H. Obayashi for his excellent presentation at the Gunma Astronomical Observatory about the first result of the M87 Event Horizon Telescope. K.A. and G.K. thank support, in part, from PREST, JST JP-MJPR1861 and from the JSPS Kakenhi (C) No. 17K05591, respectively.

\appendix
\section{Appendix}

In this Appendix, we consider practical Hawking radiation where a system $r_j$ in the thermal bath $R$ has an extremely low but nonzero temperature $\beta_{r_j}^{-1}$. In particular, here we show that even in this framework of practical Hawking radiation, Eq.~(\ref{eq:echange}) can hold when $\beta_H^{-1} \ge \beta_{r_j}^{-1}$.

We first define a two-mode squeezing operator for bosonic annihilation operators $\hat{a}_{\rm out}$ on mode $a$ and $\hat{b}_{\rm out}$ on mode $b$ as 
\begin{equation}
\hat{S}(r):=\exp [r (\hat{a}_{\rm out}^\dag \hat{b}_{\rm out}^\dag  -\hat{a}_{\rm out} \hat{b}_{\rm out})].
\end{equation}
This squeezing corresponds to a unitary inducing Hawking radiation, by taking
\begin{equation}
\tanh r = \exp[-\beta_H \omega'/2]
\end{equation}
and by regarding $a_{\rm out}$ and $b_{\rm out}$ as annihilation operators on modes for positive-heat particles $H^+$ and for negative-heat particles $H^-$, respectively.
Then, we have 
\begin{align}
&\hat{a}_{\rm in}:=\hat{S}(r) \hat{a}_{\rm out} \hat{S}^\dag(r) =\hat{a}_{\rm out} \cosh r - \hat{b}_{\rm out}^\dag \sinh r, \\
&\hat{b}_{\rm in}:=\hat{S}(r) \hat{b}_{\rm out} \hat{S}^\dag(r) =\hat{b}_{\rm out} \cosh r - \hat{a}_{\rm out}^\dag \sinh r.
\end{align}
Here $\hat{a}_{\rm in}$ and $\hat{b}_{\rm in}$ are bosonic annihilation operators on mode $a$ and mode $b$, respectively. 
If we define quadratures as
\begin{align}
\hat{q}_{c}:=&\hat{c}+\hat{c}^\dag,\\
\hat{p}_c:=& i (\hat{c}^\dag-\hat{c}),
\end{align}
for $c=a_{\rm out},b_{\rm out},a_{\rm in},b_{\rm in}$, we have
\begin{align}
\left(
    \begin{array}{c}
      \hat{q}_{a_{\rm in}} \\
      \hat{p}_{a_{\rm in}} \\
      \hat{q}_{b_{\rm in}} \\
      \hat{p}_{b_{\rm in}} \\
    \end{array}
  \right) =
\left(
    \begin{array}{cccc}
      \cosh r &0 & - \sinh r &0 \\
       0 & \cosh r & 0 & \sinh r \\
       -\sinh r & 0 & \cosh r &0  \\
      0& \sinh r &0 &\cosh r \\
    \end{array}
  \right) 
  \left(
    \begin{array}{c}
      \hat{q}_{a_{\rm out}} \\
      \hat{p}_{a_{\rm out}} \\
      \hat{q}_{b_{\rm out}} \\
      \hat{p}_{b_{\rm out}} \\
    \end{array}
  \right)  .
\end{align}
This can be rewritten as
\begin{align}
  \left(
    \begin{array}{c}
      \hat{q}_{a_{\rm out}} \\
      \hat{p}_{a_{\rm out}} \\
      \hat{q}_{b_{\rm out}} \\
      \hat{p}_{b_{\rm out}} \\
    \end{array}
  \right) =&
\left(
    \begin{array}{cccc}
      \cosh r &0 &  \sinh r &0 \\
       0 & \cosh r & 0 & -\sinh r \\
       \sinh r & 0 & \cosh r &0  \\
      0& -\sinh r &0 &\cosh r \\
    \end{array}
  \right)
\left(
    \begin{array}{c}
      \hat{q}_{a_{\rm in}} \\
      \hat{p}_{a_{\rm in}} \\
      \hat{q}_{b_{\rm in}} \\
      \hat{p}_{b_{\rm in}} \\
    \end{array}
  \right)   \nonumber \\
  =:&S(r){\bm x},
\end{align}
where ${\bm x}:=( \hat{q}_{a_{\rm in}}, \hat{p}_{a_{\rm in}}, \hat{q}_{b_{\rm in}}, \hat{p}_{b_{\rm in}})^T$. 

\begin{figure}[t]
  \begin{center}
    \includegraphics[keepaspectratio=true,height=60mm]{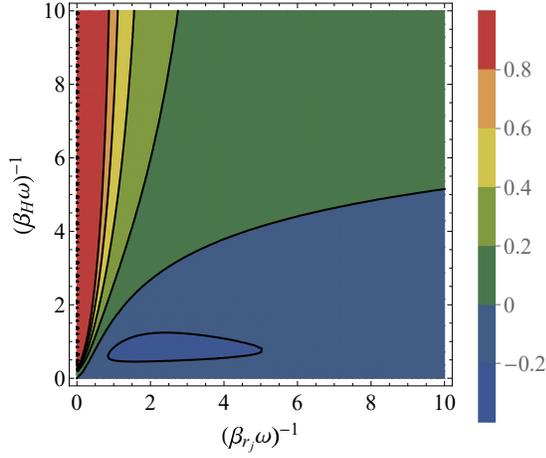}
  \end{center}
  \caption{$-\Delta_j S(B^-) +\beta_{r_j} \Delta_j {\cal Q}_{r_j}(=-\Delta S(b) +\beta_{a} \Delta {\cal Q}_{a})$. 
  From the figure, if 
the temperature $\beta_{r_j}^{-1}$ of a system $r_j$ is much lower than the Hawking temperature $\beta_H^{-1}$ like the original situation Hawking has considered \cite{H74,H75}, $-\Delta_j S(B^-) +\beta_{r_j} \Delta_j {\cal Q}_{r_j}$ can be positive, implying Eq.~(\ref{eq:echange}).}
  \label{fig:1}
\end{figure}

Now, let us assume that the mode $a$ corresponds to a system $r_j$ in the thermal bath $R$ and the initial state of mode $a$ is a Gibbs state $\hat{\chi}_{a}$ with temperature 
\begin{equation}
\beta_{a}^{-1}=\beta_{r_j}^{-1}, 
\end{equation}
while the initial state of mode $b$ is the vacuum state $\ket{\rm vac}$. In general, a Gibbs state
\begin{equation}
\hat{\chi}_a :=\sum_{m=0}^\infty \frac{n_a^m}{(n_a+1)^{m+1}} \ket{m}\bra{m}_a
\end{equation}
on mode $a$, where $n_a:=\langle \hat{a}^\dag \hat{a} \rangle={\rm Tr}[ \hat{a}^\dag \hat{a} \hat{\chi}_a]$, is completely characterized by the covariance matrix $(2n_a +1) I_{2 \times 2}$ on mode $a$ \cite{W12}, where $I_{2\times 2} := {\rm diag} (1,1)$. Therefore, the initial state of the modes $a$ and $b$ is completely characterized by the covariance matrix \cite{W12}:
\begin{align}
V_{\rm in}=&  \left(
    \begin{array}{cccc}
      2n_{a_{\rm in}}+1 &0 &  0 &0 \\
       0 & 2n_{a_{\rm in}}+1 & 0 & 0 \\
       0 & 0 & 1 &0  \\
      0& 0 &0 & 1 \\
    \end{array}
  \right) \nonumber \\
    = & \left(
    \begin{array}{cccc}
      \cosh(2s) &0 &  0 &0 \\
       0 & \cosh(2s) & 0 & 0 \\
       0 & 0 & 1 &0  \\
      0& 0 &0 & 1 \\
    \end{array}
  \right),
\end{align}
where 
$
n_{a_{\rm in}} := [\exp(\beta_a \omega') -1]^{-1}
$ and 
\begin{equation}
\tanh s=\exp(-\beta_a \omega'/2) 
\end{equation}
with a mode frequency $\omega'$. 
If the two-mode squeezing operation $\hat{S}(r)$ corresponding to Hawking radiation is applied to the modes $a$ and $b$, the final state is still a Gaussian state, which is completely characterized \cite{W12} by the covariance matrix
\begin{widetext}
\begin{align}
V_{\rm out}=S(r)V_{\rm in} S^T(r) 
  =
\left(
\begin{array}{cc}
 [\cosh (2 s) \cosh ^2 r+\sinh ^2 r] I_{2 \times 2} &  \cosh ^2 s \sinh (2 r) Z_{2\times 2}\\
  \cosh ^2 s \sinh (2 r) Z_{2\times 2} &  [ \cosh ^2 r+\cosh (2 s) \sinh ^2 r] I_{2 \times 2}\\
\end{array}
\right),
\end{align}
\end{widetext}
where $Z_{\rm 2 \times 2}={\rm diag}(1,-1)$.
From this covariance matrix, we can conclude that the final states of modes $a$ and $b$ are Gibbs states with 
\begin{align}
& 2 n_{{a}_{\rm out}} +1 = \cosh (2 s) \cosh ^2 r+\sinh ^2 r, \label{eq:nn1} \\
& 2 n_{b_{\rm out}} +1 =  \cosh ^2 r+\cosh (2 s) \sinh ^2 r, \label{eq:nn2}
\end{align}
respectively.

Finally, let us derive the analytic expression for $-\Delta S(b) +\beta_a \Delta {\cal Q}_a$, associated with Eq.~(\ref{eq:echange}). From Eqs.~(\ref{eq:nn1}) and (\ref{eq:nn2}), we have
\begin{align}
\Delta S(b):=&S(b_{\rm out}) -S(b_{\rm in}) = S(b_{\rm out}) \nonumber \\
 =& -n_{b_{\rm out}} \ln n_{b_{\rm out}} +(n_{b_{\rm out}} +1) \ln (n_{b_{\rm out}} +1) \nonumber \\
=&\frac{1}{2} \left[\sinh ^2 r \cosh (2 s)+\cosh ^2 r+1\right] \nonumber \\
& \times \ln \left[\frac{1}{2} \left(\sinh ^2 r \cosh (2
   s)+\cosh ^2 r+1\right)\right] \nonumber \\
  & -\sinh ^2 r \cosh ^2 s \ln \left[\sinh ^2 r  \cosh ^2 s \right] ,\\
\beta_a \Delta {\cal Q}_a :=& \beta_a \omega' (n_{a_{\rm out}} - n_{a_{\rm in}}) 
= \sinh ^2r \cosh ^2s \ln[\coth^2 s].
\end{align}
With these expressions, $-\Delta S(b) +\beta_a \Delta {\cal Q}_a $ is shown in Fig.~\ref{fig:1}, where we use $-\Delta S(b) +\beta_a \Delta {\cal Q}_a = -\Delta_j S(B^-) +\beta_{r_j} \Delta_j {\cal Q}_{r_j}$ by regarding mode $a$ as a system $r_j$ in the thermal bath $R$ and mode $b$ as a system in the negative-heat part $B^-$ of the black hole $B$. From the figure, if 
the temperature $\beta_{r_j}^{-1}$ of a system $r_j$ is much lower than the Hawking temperature $\beta_H^{-1}$ like the original situation Hawking has considered \cite{H74,H75}, $-\Delta_j S(B^-) +\beta_{r_j} \Delta_j {\cal Q}_{r_j}$ can be positive, that is, Eq.~(\ref{eq:echange}) can hold.

On the other hand, $\Delta S(a)$ is 
\begin{align}
\Delta S(a):=&S(a_{\rm out}) -S(a_{\rm in})  \nonumber \\
 =& -n_{a_{\rm out}} \ln n_{b_{\rm out}} +(n_{a_{\rm out}} +1) \ln [n_{a_{\rm out}} +1 ] \nonumber \\
 & +n_{a_{\rm in}} \ln n_{b_{\rm in}} -(n_{a_{\rm in}} +1) \ln [n_{a_{\rm in}} +1 ] \nonumber \\
=&\cosh ^2 s \left(\cosh ^2 r \ln \left[\cosh ^2r \cosh ^2s\right]-\ln \left[\cosh
   ^2s\right]\right) \nonumber \\
   &-\frac{1}{2} \left[\cosh ^2r \cosh (2 s)+\sinh ^2r-1\right] \nonumber \\ 
   &\times \ln
   \left[\frac{1}{2} \left(\cosh ^2r \cosh (2 s)+\sinh ^2r-1\right)\right] \nonumber \\
   &+\sinh ^2s
   \ln \left[\sinh ^2s\right].
\end{align}
Therefore, $\Delta S(b)-\Delta S(a)$ is described by
\begin{multline}
\Delta S(b)- \Delta S(a) = -\cosh ^2r \cosh ^2s \ln \left[\cosh ^2r \cosh ^2s\right]   \\
- \sinh ^2r \cosh ^2s \ln \left[\sinh ^2r \cosh ^2s\right]  \\
+\frac{1}{2} \left[\cosh ^2r \cosh (2 s)+\sinh
   ^2r-1\right]  \\ 
    \times \ln \left[\frac{1}{2} \left(\cosh ^2r \cosh (2 s)+\sinh ^2r-1\right)\right] \\
   +\frac{1}{2} \left[\sinh ^2r \cosh (2 s)+\cosh ^2r+1\right] \\
\times \ln \left[\frac{1}{2} \left(\sinh ^2r
   \cosh (2 s)+\cosh ^2r+1\right)\right]  \\
  -\sinh ^2s \ln \left[\sinh ^2s\right]+\cosh ^2s \ln \left[\cosh ^2s\right],
\end{multline}
from which we can show $\Delta S(b)- \Delta S(a) \ge 0$ analytically.

If one wants to associate the model here with 't Hooft's model~\cite{H96} for Hawking radiation, one should regard
\begin{align}
&\hat{a}_{\rm in} \to \hat{a}_2 (\tilde{k},\omega), \\
&\hat{b}_{\rm in} \to  \hat{a}_2 (-\tilde{k},- \omega), \\
& \hat{a}_{\rm out} \to \hat{a}_{\rm I} (\tilde{k},\omega) , \\
& \hat{a}_{\rm out} \to \hat{a}_{\rm II} (-\tilde{k},\omega),
\end{align}
for annihilation operators $\hat{a}_2 (\tilde{k},\omega),\hat{a}_2 (-\tilde{k},- \omega),\hat{a}_{\rm I} (\tilde{k},\omega)$ and $\hat{a}_{\rm II} (-\tilde{k},\omega)$ in Ref.~\cite{H96}. In Ref.~\cite{H96}, $\hat{a}_2$ is an annihilation operator for a freely falling observer, while $\hat{a}_{\rm I}$ and $\hat{a}_{\rm II}$ are ones on modes in regions with $\rho>0$ and with $\rho<0$ for a Rindler space coordinate $\{\tau, \rho,\tilde{x} \}$, respectively.

\end{document}